\documentclass[aps,prd,twocolumn,preprintnumbers,
superscriptaddress,showpacs,floatfix,nofootinbib]{revtex4}

\usepackage{bm}
\usepackage{epsfig}
\usepackage{graphics}

\usepackage{graphicx}

\begin{document}

\newcommand{\lsim}{\mbox{\raisebox{-.9ex}{~$\stackrel{\mbox{$<$}}{\sim}$~}}}
\newcommand{\gsim}{\mbox{\raisebox{-.9ex}{~$\stackrel{\mbox{$>$}}{\sim}$~}}}

\title{Cosmic Superstrings and Primordial Magnetogenesis}

\author{Anne--Christine Davis}
\email{acd@damtp.cam.ac.uk}
\affiliation{Department of Applied Mathematics and Theoretical Physics, 
Centre for Mathematical Sciences, Wilberforce Road, Cambridge CB3 0WA, UK}
\author{Konstantinos Dimopoulos}
\email{k.dimopoulos1@lancaster.ac.uk}
\affiliation{Department of Physics, Lancaster University, Lancaster LA1 4YB,
 UK}

\date{\today}

\begin{abstract} 
Cosmic superstrings are produced at the end of brane inflation. Their 
properties are similar to cosmic strings arising in grand unified
theories. Like cosmic strings they can give rise to a primordial magnetic
field, as a result of vortical motions stirred in the ionised plasma
by the gravitational pull of moving string segments. The resulting magnetic 
field is both strong enough and coherent enough to seed the galactic dynamo
and explain the observed magnetic fields of the galaxies. 
\end{abstract}

\maketitle

\section{Introduction}
Cosmic superstrings have received a lot of interest recently due to 
developments in fundamental string theory. They arise
naturally in models of brane inflation and have characteristic 
differences with ordinary cosmic strings, which could provide
a window into string theory. There is also the distinct possibility
that they could solve some long-standing problems in cosmology
and astroparticle physics.

For example, magnetic fields pervade most astrophysical 
objects \cite{kron,beck}, but their origin is still elusive. In the last decade
a number of attempts were made to explain the observed magnetic fields of the 
galaxies, none of which has been conclusive. Many authors have considered that 
these magnetic fields originate from the Early Universe and are truly 
primordial. Since a large scale primordial magnetic field
(PMF) cannot be generated in thermal equilibrium (because it 
breaks isotropy), research has been focused in magnetogenesis mechanisms 
either during phase transitions or from inflation. Phase transitions occur 
very early in the history of the Universe. Consequently, any PMF generation
creates highly incoherent magnetic fields \cite{P/T} that cannot give rise to 
the magnetic fields of the galaxies (unless one considers inverse cascade 
mechanisms \cite{inverse}). On the other hand, due to the conformal invariance 
of electromagnetism, generating a PMF during inflation substantially
dilutes its strength down to insignificant values \cite{TW} (see however 
\cite{NM}). An extensive review of the literature on PMFs can be found in 
\cite{pmf} (see also references in \cite{NM}).

Recent developments in string theory offer another possibility that PMFs 
could have a fundamental origin. 
The motion of a network of cosmic strings can result in a primordial magnetic
field which is strong enough to seed the galactic dynamo 
\cite{VV,vacha,voll,AS,mine}. However, 
cosmic strings arising in grand unified theories seem to be at variance
with the observations of the CMB. Models of brane inflation predict
the formation of cosmic superstrings at the end of inflation
\cite{tye,copeland}. Such
cosmic superstrings have a lower string tension than those arising
in grand unified theories, so they evade the CMB limits on cosmic strings
(for reviews see \cite{polchinski,davis&kibble}).
Consequently, a network of cosmic superstrings could produce a 
viable primordial magnetic field and still be consistent with other
cosmological observations. 

In this paper we investigate this possibility, presenting two mechanisms for
the production of a primordial magnetic field from a network of 
cosmic superstrings. In Sec.~II we discuss cosmic superstrings
and their characteristics; their tension and intercommutation probability.
In Sec.~III we present the magnetogenesis mechanism, based on the effect of a 
network of cosmic superstrings onto ionised plasma. We consider two 
realisations of this mechanism; one generating a PMF inside the string wakes 
and the other over inter-string distances. We also consider the cases of wiggly
strings or current carrying strings. Finally, in Sec.~IV we discuss our results
and present our conclusions. Throughout the paper we use natural units, 
such that \mbox{$c=\hbar=1$}. The signature of the spacetime metric is taken to
be $(-,+,+,+)$.

\section{Cosmic Superstrings}
There has been a resurgence of interest in cosmic strings arising
from recent results in fundamental string theory. Indeed, they are predicted
to arise in models of brane inflation where an extra brane and anti-brane 
annihilate to produce lower dimensional branes, with the inter-brane 
distance playing the role of the inflaton. In this picture D-strings, 
or $D1$ branes are formed generically \cite{tye}. Similarly fundamental 
strings, or F-strings, can also arise \cite{copeland} and, in certain
classes of models, axionic local strings \cite{DBD}. 

In models of brane inflation the extra brane and anti-brane are localised at 
the bottom of a throat in the compact dimensions. Consquentially, 
D-strings and F-strings are also formed in the throat. Since space-time is
highly warped in the throat this results in
the string tension of the D- and F-strings being less than the fundamental
scale,
\begin{equation}
\mu=e^{-A(y)}\mu_0,
\end{equation}
where $A(y)$ is the warp factor with $y$ refering to the compact dimensions
and $\mu_0$ is the fundamental scale.
Estimates give the range to be between
$10^{-12}\leq G\mu\leq 10^{-6}$ depending on details  of the
theory (see \cite{polchinski, davis&kibble} for a review). 

The gravitational effects of cosmic superstrings will be similar to
those of the usual cosmic strings and they will be subject to the
same constraints. For example, we know that cosmic strings are not
the primary source of structure formation, resulting in a constraint
of $G\mu<3.3\times10^{-7}$ arising from the WMAP data \cite{smoot}. Similary
the regularity of the pulsar timings results in a constraint on 
gravitational waves emitted by cosmic strings, corresponding to
$G\mu<10^{-7}$ (see for example \cite{tom} and references therein). 
Hence we arrive at the range
\begin{equation}
10^{-12}\leq G\mu<10^{-7}
\label{Gmu}
\end{equation}

Supersymmetric theories give rise to two sorts
of strings, called  D-term or F-term strings \cite{DDT}, where
the D and F refer to the  type of potential required to break
the symmetry. A natural  question to ask is whether these
cosmic strings are related to the D- and F-strings discussed
above. A recent analysis of  supersymmetric theories with a
D-term suggests that D-term cosmic strings may well be
D-strings \cite{dvali}. It is then possible that D-strings
are current-carrying via fermion zero modes since it was shown
that fermion zero modes survive supersymmetry breaking for
D-term strings \cite{DDT2}. 

The cosmology of D-strings (and F-strings) is a little different
from that of ordinary cosmic strings. For ordinary cosmic strings, the
probability of intercommutation is $P\simeq 1$. This is not the case
for D-strings since they can `miss'  each other in the compact dimension,
whilst for F-strings intercommutation is a quantum mechanical process. 
The probability of intercommuting has  been estimated to be
between $10^{-1}\leq P\leq 1$ for D-strings and $10^{-3}\leq P\leq 1$ for 
F-strings \cite{polchinski}. Similarly the probability of a string
self-intersecting is reduced. This means that a network of
such strings  could look different from that of cosmic
strings. There are suggestions that such a network would be
denser, with the distance between strings related to $P$, and
slower \cite{dvali&vilenkin, mairi}.  It is likely
that the net result would be  to increase the number of string
loops, despite the reduction in string self-intersection. A
network of D-strings could also emit exotic particles, such as
dilatons \cite{DV2,babichev}, as a result of the underlying superstring 
theory.  

The evolution of cosmic superstrings will vary from that of cosmic
strings. Usually a cosmic string network reaches a scaling solution.
For cosmic superstrings this is still the case however the intercommutation
probability comes into the scaling solution. There have been analytic 
\cite{dvali&vilenkin, avgoustidis&shellard} and
numerical \cite{mairi} investigations into the behaviour of the cosmic 
superstring network, leading to the conclusion that the correlation 
length behaves as
\begin{equation}
\xi=P^{\beta}t,\quad{\rm where}\quad\frac{1}{2}\leq\beta\leq 1
\label{xi}
\end{equation}
Similarly the gravitational
radiation emitted from a cosmic superstring network will depend on
the parameter $P$, which losens the pulsar constraints on $G\mu$ discussed
above.

Cosmic strings can generate a primordial magnetic field
\cite{VV,vacha,voll,AS,mine}.
Similarly we would expect D-strings to give rise to a primordial 
magnetic field in a similar way to other local cosmic strings.
However, there will be distinct differences for D-strings given that 
their cosmology differs. In some models semi-local strings arise 
\cite{jon,kallosh}, rather than cosmic strings. These are not topologically
stable \cite{ana&tanmay}, but if they were to live long, they could still
contribute in a similar way to cosmic strings, and similarly for the
local axionic strings \cite{DBD}.

For F-strings there could still be a primordial magnetic field produced
due to the motion of the string through the surrounding plasma. Here,
though, the mechanism will be similar to that for global strings.

In the next section we review magnetogenesis mechanisms with cosmic string
networks.

\section{Magnetogenesis mechanisms}

In this section we will study two mechanisms for the generation of a 
primordial magnetic field (PMF) due to the cosmological effects of a network
of cosmic superstrings. 
%
Although the existence of a PMF may have many cosmological implications we
will focus more on the possibility of explaining the galactic magnetic fields,
by triggering the $\alpha-\Omega$ dynamo mechanism in galaxies after galaxy 
formation. Such a mechanism requires the presence of a preexisting seed 
magnetic field in order to operate. This seed field has to satisfy certain 
requirements in terms of strength and coherence. ~These are the following.

To successfully trigger the dynamo and explain the galactic magnetic 
fields the lower bound on the strength of the seed field (in a dark energy 
dominated Universe) is \cite{anne}
\begin{equation}
B_{\rm seed}\geq 10^{-30}{\rm Gauss}
\label{Bbound}
\end{equation}
Such a seed field is amplified exponentially by the galactic dynamo until it 
reaches the observed value \mbox{$B_{\rm obs}\sim 10^{-6}$Gauss}, where it
becomes dynamically important (its energy is comparable to the kinetic energy 
of galactic rotation). At this stage galactic dynamics backreacts to the dynamo
mechanism and stabilises the value of the field. Considering the characteristic
timescale for the dynamo operation (the galactic rotation period) a seed field 
weaker that the bound in Eq.~(\ref{Bbound}) would not have enough time to be
amplified up to the observed value.\footnote{%
Before the discovery of dark energy the lower bound on $B_{\rm seed}$ was much 
more stringent: \mbox{$B_{\rm seed}\geq 10^{-21}$Gauss}. This is easy to 
understand as follows. The minimum strength of the seed field corresponds to a 
field which, when amplified by the dynamo from the time of galaxy formation 
$t_{\rm gf}$ until the present time $t_0$, just about reaches the observed 
value of 1~$\mu$Gauss. Hence, we have 
\mbox{$1~\mu{\rm Gauss}\sim e^NB_{\rm seed}^{\rm min}$}, where $N$ is the 
number of galactic revolutions since the time of galaxy formation. Now, 
\mbox{$N\sim\Delta t/\tau_g$}, where \mbox{$\Delta t=t_0-t_{\rm gf}\simeq t_0$}
and $\tau_g$ is the timescale of dynamo amplification (galactic rotation 
period). Without dark energy
\mbox{$t_0\simeq 8.96$ Gyrs}, which suggests that the galaxy has rotated about
\mbox{$N\simeq 35$} times. However, when taking dark energy into account, the
age of the Universe is multiplied by a factor 
\mbox{$\frac{1}{\sqrt{\Omega_\Lambda}}$%
sinh$^{-1}\sqrt{\frac{\Omega_\Lambda}{1-\Omega_\Lambda}}$}, which, for
\mbox{$\Omega_\Lambda\simeq 0.7$}, gives \mbox{$t_0'\simeq 13.7$ Gyrs}. 
Thus, the number of galactic revolutions becomes
\mbox{$N'=(13.7/8.96)N\simeq 54$}. Hence, the lower bound on the seed field
now reeds: \mbox{$B_{\rm seed}^{\rm min}\sim e^{-N'}\times1\;\mu$Gauss
$\sim 10^{-30}$Gauss}.}

Also, for the dynamo action not to be destabilised the seed field has to 
avoid being too incoherent. Indeed the coherence of the seed field cannot be
much smaller than~\cite{ander}
\begin{equation}
\ell_{\rm seed}\gsim 100\;{\rm pc}\,.
\label{lbound}
\end{equation}

In both the mechanisms that we consider the PMF generation is based on the 
Harrison--Rees mechanism, which is briefly reviewed below.

\subsection{The Harrison--Rees mechanism}

Harrison was the first to consider the generation of a magnetic field by the 
vortical motions of ionised plasma. He suggested that turbulence in an 
expanding Universe may generate a magnetic field since the turbulent velocity 
would be different for the electrons and the much heavier ions \cite{hary}. 
His argument focused in the radiation era and can be sketched as follows.

Consider a rotating volume $V$ of ionised plasma. Suppose that the angular
velocities $\omega_i$ and $\omega_e$ of the ion and the electron fluid 
respectively are uniform inside $V$. Then, since \mbox{$V\propto a^3$}, we
find that
\begin{equation}
\rho_i V={\rm const.}\qquad{\rm and}\qquad\rho_e V^{4/3}={\rm const.}\;,
\end{equation}
where \mbox{$\rho_i\propto a^{-3}$} is the ion density, which scales like 
pressureless matter, while \mbox{$\rho_e\propto a^{-4}$} is the electron
density, which scales as radiation due to the strong coupling between the 
electrons and the photons, through Thompson scattering. The angular momentum
\mbox{${\cal I}=\rho\omega V^{5/3}$} of each plasma component has to be 
conserved. This suggests that 
\begin{equation}
\omega_i\propto V^{-2/3}\propto a^{-2}\quad{\rm and}\quad
\omega_e\propto V^{-1/3}\propto a^{-1}.
\end{equation}

Thus, the ion fluid spins down faster than the electron-photon gas. 
Consequently, a circular current is generated, which creates a magnetic field 
in the volume $V$.

Rees, however, has shown that expanding volumes of spinning plasma are unstable
in the radiation era and decay with cosmic expansion 
\cite{rees}. He suggested instead a different version of vortical magnetic 
field generation involving Compton scattering of the electrons on the CMB
(Compton drag mechanism). This applies after recombination and tends to damp 
the vortical motions of the electrons in contrast to ions, which remain 
unaffected. The result is again the generation of circular currents but, this 
time, it is the electron fluid that slows down. 

In both cases, the Maxwell's equations suggest \cite{hary}
\begin{equation}
{\bf B}\simeq-\frac{m_p}{e}{\bf w}\,,
\label{Brot}
\end{equation}
where \mbox{$m_p\sim 1$ GeV} is the nucleon mass and ${\bf w}$ is the vorticity
of the plasma, given by
\begin{equation}
{\bf w}=\nabla\times{\bf v}_{\rm rot}\;,
\label{w}
\end{equation}
with ${\bf v}_{\rm rot}$ being the rotational velocity of the spinning plasma.
A similar mechanism is presented in \cite{misruz}.

\subsection{Vortical motions inside the string wakes}

Vachaspati and Vilenkin were the first to suggest that vortical motions inside 
the wakes of cosmic strings can give rise to PMFs \cite{VV,vacha} (see also 
\cite{voll}). The idea is that the boost generated by the deficit angle of the 
cosmic string metric may stir vorticity in the matter, which falls into the 
wake of a travelling string. The vortical motions themselves are 
generated by the rapidly changing conical metric of the string in the 
small-scale wiggles, which a long string develops due to self-intersections. 
The oscillations of the wiggles are expected to generate turbulence in the 
plasma inside the string wake.

The metric of the (2+1)-dimensional spacetime perpendicular to a straight 
string 
is:
\begin{equation}
ds_\perp^2=-dt^2
+dr^2+(1-8G\mu)r^2d\phi^2,
\label{metric}
\end{equation}
which describes the space around a cosmic string as Euclidean with a 
wedge of angular size $\Delta$ removed, where 
\begin{equation}
\Delta=8\pi G\mu\,.
\label{D}
\end{equation}
A test particle at 
rest with respect to the string experiences no gravitational force but, if the 
string moves with velocity $v_s$, then nearby matter undergoes a boost
\begin{equation}
u=4\pi G\mu v_s\gamma_s\;
\label{u}
\end{equation}
in the direction perpendicular to the motion of the string, where 
\mbox{$\gamma_s=1/\sqrt{1-v_s^2}$}. The above boost is the characteristic 
velocity of the turbulence caused by the wiggles, i.e. 
\mbox{$v_{\rm rot}\simeq u$}. Hence, an estimate of the vorticity is
\begin{equation}
|{\bf w}|\simeq\frac{v_{\rm rot}}{R}\simeq\frac{4\pi}{\Gamma t}\,,
\label{w1}
\end{equation}
where we used that the characteristic length-scale of the wiggles is given by
\begin{equation}
R\simeq\Gamma G\mu t\,,
\label{G}
\end{equation}
with $\Gamma$ determined by the rate of emission of gravitational radiation 
from the string, due to the oscillating wiggles. For gauge strings, 
simulations have shown that \mbox{$\Gamma\sim 100$}. For cosmic superstrings 
this may change somewhat because the intercommutation of the wiggles is 
suppressed and therefore one has less efficient loop formation and less kinks 
on the long string. Also, gravitational radiation may escape in the 
extra dimensions, though detailed simulations have yet to be performed
\cite{damour&vilenkin}.

Using the above, Eq.~(\ref{w1}) suggests that the PMF generated at some time 
$t_f$ is 
\begin{equation}
B_f
\sim\frac{m_p}{e}\frac{4\pi}{\Gamma t_f}\,.
\end{equation}
Note that, remarkably, $B_f$ does not depend on the value of $G\mu$.

Due to the high conductivity of the plasma the PMF is expected to freeze onto 
the plasma. The conservative approach, then, is to consider that the turbulent 
eddy is not gravitationally bound. This is reasonable to expect because cosmic 
string wake formation is no longer associated with structure formation, the 
latter occurring at overdensities generated due to inflation, which dominate 
the wake overdensities. For a non-gravitationally bound eddy one may estimate 
the strength of the magnetic field at galaxy formation by assuming that, being 
frozen into the plasma, the magnetic field conserves its flux and, therefore, 
scales as \mbox{$B\propto a^{-2}$}. Thus, scaling the above PMF down to
the time of galaxy formation we obtain
\begin{equation}
B_{\rm gf}\sim 
B_f
\left(\frac{a_f}{a_{\rm gf}}\right)^2\sim
\frac{4\pi m_p}{e\Gamma t_f}
\left(\frac{t_f}{t_0}\right)^{4/3}\!
(z_{\rm gf}+1)^2,
\label{Bgf}
\end{equation}
where $t_0$ is the present time, \mbox{$z_{\rm gf}\sim 6$} is the redshift at
galaxy formation and we took into account that, until galaxy formation, the 
Universe remains matter dominated, i.e. \mbox{$a\propto t^{2/3}$}. 

Similarly, we can find the coherence length of the magnetic field at galaxy 
formation. At formation the coherence scale is determined by the scale of the 
wiggles which stir the vortical motion [cf. Eq.~(\ref{G})]
\begin{equation}
\ell_f\sim\Gamma G\mu t_f\;.
\end{equation}
Since we consider an eddy which is not gravitationally bound 
\mbox{$\ell\propto a$}. Hence, at galaxy formation we find
\begin{equation}
\ell_{\rm gf}\sim\left(\frac{a_{\rm gf}}{a_f}\right)\ell_f\sim
\frac{\Gamma G\mu t_f}{(z_{\rm gf}+1)}
\left(\frac{t_0}{t_f}\right)^{2/3}
\label{lgf}
\end{equation}

Eqs.~(\ref{Bgf}) and (\ref{lgf}) show that the dependence of the PMF
strength and coherence on the time of formation $t_f$ is very weak:
\begin{equation}
B_{\rm gf}, \ell_{\rm gf}\propto t_f^{1/3}
\label{t1/3}
\end{equation}
Indeed, it can be easily checked that, for $t_f$ between recombination and 
galaxy formation, the variance of both these quantities is no more than an
order of magnitude with the best results achieved when the PMF is generated at 
late times. Hence, adopting again a conservative approach we estimate 
$B_{\rm gf}$ and $\ell_{\rm gf}$ at the time of recombination $t_{\rm rec}$.

After recombination there is some residual ionisation present in the plasma, 
which can allow the vortical generation of PMFs \cite{misruz}. Setting
\mbox{$t_f=t_{\rm rec}$}, it is easy to find
\begin{equation}
B_{\rm gf}\sim 10^{-23}{\rm Gauss}\quad{\rm and}\quad
\ell_{\rm gf}\sim 10^4(G\mu)\,{\rm Mpc}\,,
\end{equation}
where we used \mbox{$v_s\gamma_s\sim 1$} and \mbox{$\Gamma\sim 100$}. 
If this PMF is carried by the plasma during the gravitational collapse of
a galaxy, then flux conservation amplifies its strength by a factor
$$\left(\frac{{\rm intergalactic\;distance\;at}\;
t_{\rm gf}}{{\rm galactic\;size}}\right)^2\sim{\cal O}(10^2)$$
while its coherence is decreased by a factor
$$\frac{{\rm galactic\;size}}{{\rm intergalactic\;distance\;at}\;t_{\rm gf}}
\sim{\cal O}(10^{-1})\,.$$ Hence, the seed field for the galactic dynamo is
\begin{equation}
B_{\rm seed}\sim 10^{-21}{\rm Gauss}\quad{\rm and}\quad
\ell_{\rm seed}\sim 10^3(G\mu)\,{\rm Mpc}\,.
\label{Blseed}
\end{equation}
Comparing the above with the bound in Eq.~(\ref{Bbound}) we see that
such a seed field is strong enough to successfully trigger the dynamo and 
explain the galactic magnetic fields. However, the coherence requirements are
more difficult to satisfy. Indeed, comparing the above to the bound in 
Eq.~(\ref{lbound}) we see that the latter can be satisfied only if 
\mbox{$G\mu\gsim 10^{-7}$}, which is in marginal conflict with the 
observations.

The situation can be somewhat improved if we consider PMF generation at much
later times than recombination. The latest appropriate time corresponds to
the epoch of earlier ionisation that precedes galaxy formation.
Reionisation of the Universe at late times has indeed been detected by the 
WMAP, based on the observed decrease of the temperature angular power spectrum 
at high multiples and by an excess in the TE cross-power spectrum on large 
angular scales, with respect to the case of no or little reionisation. Many 
believe that this reionisation occurs at two stages; late reionisation due to 
quasars at redshifts \mbox{$z_{\rm ri}\simeq 6$} and earlier reionisation at
redshifts of at least \mbox{$z_{\rm ri}\gsim 15$} and up to (a few)$\times$10
possibly associated to Population~III stars \cite{wmap}. 

Taking \mbox{$z_f=z_{\rm ri}\simeq 15$} it is easy to find that, due to 
Eq.~(\ref{t1/3}), both strength and coherence of the seed field are 
intensified by an order of magnitude. However, generating the PMF that late 
implies that only a small fraction of the galaxies can benefit from the
mechanism. This is because, even though the low intercommutation probability 
$P$ results in a denser string network, one cannot envisage more than about 
\mbox{$t/\xi\leq P^{-1}\lsim 10^3$} [cf. Eq.~(\ref{xi})]
long strings travelling at the comoving volume of the present horizon at the 
time of formation of the PMF. This means that magnetisation will appear in 
thin sheets of width given roughly by $t_f^{-1}$, which, for 
\mbox{$z_f=z_{\rm ri}$} could be quite far apart (comoving distance: 
$\sim$~100 Mpc), leaving a lot of the protogalaxies ``untouched''. Also, since 
structure formation is not really related to string wakes, it is not certain 
how much of the magnetised plasma will find its way into galaxies (even though 
magnetised plasma dispenses more efficiently with angular momentum, which 
assists gravitational collapse). A fair portion of magnetised plasma will 
remain in the intergalactic medium and will result in intergalactic magnetic 
fields of order $10^{-24}$Gauss, which are far weaker than the ones observed 
\cite{kron} (the latter are thought to be expelled to the IGM by active 
galaxies through processes such as the Parker instability). 

In contrast, a PMF generated just after recombination permeates most of the 
plasma because, at recombination the string network will be denser by a factor
$$\left(\frac{t_{\rm ri}}{t_{\rm rec}}\right)\times
\left(\frac{a_{\rm rec}}{a_{\rm ri}}\right)\sim
\left(\frac{t_{\rm ri}}{t_{\rm rec}}\right)^{1/3}\sim{\cal O}(10^2)$$
assuming it follows a scaling solution with $P^{-\beta}$ strings per horizon. 
Hence the magnetised sheets 
could be as close as \mbox{$\sim$ 1~Mpc} comoving distance.

Still, stability arguments may inhibit the generation of a PMF for 
non-gravitationally bound eddies \cite{rees}. If this is so then we have to
limit ourselves to gravitationally bound objects that have been captured by
the rapidly oscillating wiggles, while the string traverses space. This would 
imply a much stronger PMF since there will be no further dilution due to the 
expansion of the Universe. However, this also means that not all the turbulent 
material in the string wake can be expected to become magnetised but only any 
preexisting lumps that have been caught by the passage of the string. 
Furthermore, since the dimensions of the magnetised region would not follow the
expansion of the Universe, the coherence of the PMF would be much less than 
previously considered because \mbox{$\ell_{\rm gf}\simeq\ell_f$}. This allows 
the possibility to satisfy the bound in Eq.~(\ref{lbound}) only if the PMF is 
generated rather late, i.e. at the reionisation time. However, as we have 
already mentioned, at late times the string wakes are far apart, which means
that galaxies would only be sparsely magnetised.


\subsection{Vortical motions on inter-string distances}

An alternative way for the generation of a PMF by a network of cosmic strings
through the Harrison--Rees mechanism is considering vortical motions in the 
plasma stirred by travelling neighbouring strings in a string network.
Travelling cosmic strings can also drag the plasma behind them. 
This causes circular motions over the inter-string distance (the separation
between two neighbouring strings in the network) as neighbouring strings pass 
by one another in opposite directions. 

One important aspect of the mechanism is the consideration of cosmic strings 
which also develop an {\em attractive} gravitational field, which assists to 
the drag of the plasma in the string trail. In order for this to occur we have 
to consider strings whose metric is slightly different compared to 
Eq.~(\ref{metric}).

Let us consider a straight cosmic string, whose (2+1)-dimensional perpendicular
spacetime is:
\begin{equation}
ds_\perp^2=(1-h_{00})[-dt^2+dr^2+(1-\Delta/\pi)r^2d\phi^2],
\label{metric1}
\end{equation}
where $h_{00}$ is the time-time component of a perturbation of the metric
\mbox{$g_{\mu\nu}=\eta_{\mu\nu}+h_{\mu\nu}$}, with $\eta_{\mu\nu}$ being the 
metric of Minkowski spacetime. $h_{00}$ can be non-zero in models, where the 
effective energy per unit length $\tilde\mu$ is different that the tension $T$ 
of the string. From the above we see that spacetime remains conical, with 
a deficit angle $\Delta$. However, due to the \mbox{$(1-h_{00})$} factor
there is also some gravitational attractive force towards the string. This can 
be seen as follows.

The geodesic equation is, \mbox{$\frac{d^2u^\lambda}{d\tau^2}+
\Gamma^\lambda_{\mu\nu}u^\mu u^\nu=0$}, where 
\mbox{$u^\mu=dx^\mu/d\tau=(1,{\bf v})$} is the 4-velocity and 
\mbox{$\Gamma^\lambda_{\mu\nu}\simeq\frac{1}{2}\eta^{\lambda\rho}
(\partial_\nu h_{\mu\rho}+\partial_\mu h_{\nu\rho}-\partial_\rho h_{\mu\nu})$} 
are the Christoffel symbols. Since, for the plasma, \mbox{$|{\bf v}|\ll 1$}, 
the geodesic equation becomes
\begin{equation}
\frac{d^2x^i}{d\tau^2}+\Gamma^i_{00}=0\,,
\label{geo}
\end{equation}
where $i$ denotes the spatial coordinates and $\tau$ 
is the proper time. Since \mbox{$\Gamma^i_{00}=-\frac{1}{2}\partial_ih_{00}$} 
we find that the gravitational force per unit length is
\begin{equation}
{\bf f}=\frac{1}{2}\nabla h_{00}\,.
\label{f}
\end{equation}

Now, let us investigate what this implies for the plasma particles when 
a string with such a gravitational field passes by. For this it is better to 
rewrite the metric in Eq.~(\ref{metric1}) in Cartesian coordinates
\begin{equation}
ds_\perp^2=(1-h_{00})(-dt^2+dx_kdx^k),
\label{metric2}
\end{equation}
where \mbox{$k=1,2$} and we need to extract from the above a wedge of deficit
angle $\Delta$. Then, we have
\begin{equation}
d\tau^2=-ds_\perp^2=(1-h_{00})dt^2(1-\dot x_k\dot x^k)\,.
\end{equation}
Using \mbox{$h_{00}$} and also that 
\mbox{$\Gamma^i_{00}=-\frac{1}{2}\partial_ih_{00}$} we insert the above into
Eq.~(\ref{geo}) and obtain
\begin{equation}
2\ddot x^i=(1-\dot x_k\dot x^k)\partial^ih_{00}\;,
\label{accel}
\end{equation}
with \mbox{$i=1,2$}. The above gives the acceleration felt by the particles due
to the gravitational pull of the string, in the frame of the string.

Suppose that the string moves in the $x$-direction with constant velocity
\mbox{$-v_s=-\sqrt{\dot x_k\dot x^k}$}. Then, for a particle in the position
$(x,y)$, we have initially \mbox{$x=vt$} and \mbox{$y=$ const.} The velocity
boost felt by the particle towards the $y$-direction after its encounter with 
the string is
\begin{equation}
u_y=\int^\infty_{-\infty}\ddot ydt=\frac{1}{2v_s\gamma_s^2}
\int^\infty_{-\infty}\partial_yh_{00}dx\,,
\end{equation}
where we also used Eq.~(\ref{accel}). Taking into account the deficit angle,
the relative boost between the particles on the opposite sides of the string is
\mbox{$\delta u_y=v_s\Delta+2|u_y|$}. Switching to the particle frame gives
\begin{equation}
u=\gamma_s\delta u_y=\Delta v_s\gamma_s+\frac{2I}{v_s\gamma_s}\,,
\label{u1}
\end{equation}
where 
\begin{equation}
I\equiv\int^\infty_{-\infty}
f_ydx\,,
\label{I}
\end{equation}
and we also considered Eq.~(\ref{f}).
In Eq.~(\ref{u1}) the first term is due to the conical spacetime [cf. 
Eqs.~(\ref{D}) and (\ref{u})], while the second term is due to the 
gravitational attractive force.

The deflection of particles in the string spacetime results in a net drag of 
the plasma behind the string. This is due to the fact that the magnitude of the
particle velocity is not modified after the interaction with the string. The 
velocity of plasma dragging can be estimated by Taylor expanding the
particle velocity given in Eq.~(\ref{u1}). To the lowest order we find
\begin{equation}
\delta v\simeq\frac{1}{2}\frac{u^2}{v_s}\,.
\label{dv}
\end{equation}

The backreaction of this effect is a decelerating force on the string,
which can be estimated as follows
\begin{equation}
f_{\rm drag}=\int\frac{d^2p}{dtdz}dxdy
\simeq 2R\rho v_s\delta v\,,
\end{equation}
where $f_{\rm drag}$ is the drag force per unit length, \mbox{$p\sim\rho v_s$} 
is the momentum of the plasma in the string frame, 
\mbox{$dx\simeq(\delta v)dt$} is the drag of the plasma and $R$ is the 
inter-string distance (over which $dy$ is integrated). 

A string segment of length $R$ may transfer
momentum to the plasma in the inter-string volume $\sim R^3$. In this way the
string network can induce vortical motions to the plasma on inter-string 
scales.
The total force
on a plasma volume of dimensions comparable to the inter-string distance $R$ is
\begin{equation}
F\simeq\int_0^Rf_{\rm drag}dz\sim R^2\rho u^2,
\label{F}
\end{equation}
where we also used Eq.~(\ref{dv}). 
Hence, the typical rotational velocity is estimated as
\begin{equation}
FR\simeq\frac{1}{2}M_Rv_{\rm rot}^2\quad\Rightarrow\quad
v_{\rm rot}\sim\frac{\sqrt{F/\rho}}{R}\sim u\,,
\end{equation}
where \mbox{$M_R\sim\rho R^3$} is the mass in the inter-string volume and we 
used Eq.~(\ref{F}).

The strength of the PMF generated by the vortical motions can be estimated 
using Eqs.~(\ref{Brot}) and (\ref{w}), with \mbox{$v_{\rm rot}\sim u$}. It
is straightforward to obtain
\begin{eqnarray}
B & \simeq & \frac{m_p}{e}\frac{v_{\rm rot}}{R}\;\;\Rightarrow\nonumber\\
B_f & \sim & \frac{m_p}{e}\frac{\gamma_s}{P^{\beta}t_f}
\left[\Delta+\frac{2I}{(v_s\gamma_s)^2}\right],
\label{Bf}
\end{eqnarray}
where \mbox{$R\sim P^{\beta}(v_st)$}.
Obviously, the coherence of the PMF is given by the inter-string 
distance, i.e.
\begin{equation}
\ell_f\sim P^{\beta}v_st_f\,.
\label{lf}
\end{equation}

In contrast to the previous case, we will concentrate on gravitationally bound
eddies, which do not suffer from stability problems. The reason is that, since
the total of inter-string volumes spans all space (in contrast to wakes behind
the moving strings), it follows that all the gravitationally bound objects lie
inside volumes that may well be rotated by the string network. From 
Eq.~(\ref{lf}) we see that the inter-string distance can be much smaller than 
the horizon and, therefore, overdensities that become causally connected, 
collapse and detach from Hubble expansion are quite likely to be affected 
by strings. We expect the most prominent PMF generation to occur near 
recombination, when the plasma is still substantially ionised and just after
structure formation begins.

In the following we will estimate the strength and coherence of the 
inter-string PMF in two cases, which can be described by a metric of the form 
shown in Eq.~(\ref{metric1}).

\subsubsection{Wiggly strings}

PMF generation on inter-string distances by a network of cosmic strings was 
first considered by Avelino and Shellard \cite{AS}. They considered the case
of wiggly strings, which can develop an attractive gravitational field because
their effective energy per unit length $\tilde\mu$ and their tension $T$ are 
different (whereas for a straight string \mbox{$T=\mu$}). 
This is due to coarse--graining the small scale structure (wiggles) 
on the string. The relation between $\tilde\mu$ and $T$ is
\begin{equation}
\tilde\mu T=\mu^2.
\end{equation}
Simulations for gauge strings
estimate \mbox{$\tilde\mu\approx 1.6\mu$}, which means that
\mbox{$\tilde\mu-T\approx 0.6\mu$}. 

For wiggly strings it has been found that \cite{VV}
\begin{equation}
h_{00}=-4G(\tilde\mu-T)\ln(r/r_0)\,,
\label{h00}
\end{equation}
where $r_0$ is the radius of the string core. Inserting this into 
Eq.~(\ref{f}) we find an attractive force per unit length
\begin{equation}
f=-\frac{2G(\tilde\mu-T)}{r}\,.
\end{equation}
Hence, the total boost is
\begin{equation}
u=8\pi G\tilde\mu v_s\gamma_s+\frac{4\pi G(\tilde\mu-T)}{v_s\gamma_s}\,,
\label{u2}
\end{equation}
where we used Eqs.~(\ref{u}) and (\ref{I}) considering also that the
deficit angle is still given by Eqs.~(\ref{D}) with 
\mbox{$\mu\rightarrow\tilde\mu$}. Using the above, Eq.~(\ref{Bf}) suggests

\begin{equation}
B_f\sim\frac{m_p}{e}\frac{8\pi G\tilde\mu\gamma_s}{P^{\beta}t_f}
\left[1+\frac{\tilde\mu-T}{2\tilde\mu(v_s\gamma_s)^2}\right].
\label{Bf1}
\end{equation}
%
Evaluating the above at recombination and considering also that the 
gravitational collapse of a galaxy amplifies the PMF by a factor of 
${\cal O}(10^2)$ we find
\begin{equation}
B_{\rm seed}\sim 10^{-15}P^{-\beta}(G\mu)\,{\rm Gauss}\,,
\label{Bseed1}
\end{equation}
where we considered \mbox{$v_s\gamma_s\sim 1$}.
Comparing the above with the bound in Eq.~(\ref{Bbound}) we see that the
generated PMF is strong enough to seed the galactic dynamo provided
\begin{equation}
G\mu\geq 10^{-15}P^{\beta},
\end{equation}
which is satisfied for the entire range of $G\mu$ shown in Eq.~(\ref{Gmu}).

With respect to coherence, evaluating Eq.~(\ref{lf}) at recombination and 
considering also that galactic gravitational collapse reduces the coherence of 
a PMF by a factor of ${\cal O}(10^{-1})$, we find
\begin{equation}
\ell_{\rm seed}\sim 10^{-2}P^{\beta}\,{\rm Mpc}\,.
\label{lseed1}
\end{equation}
Comparing this with the bound in Eq.~(\ref{lbound}) we see that the PMF is 
coherent enough for the dynamo, provided
\begin{equation}
P^{\beta}\geq 10^{-2}.
\end{equation}
From Eqs.~(\ref{Bf}) and (\ref{lf}) it is evident that if we consider later 
times for the PMF generation, the coherence of the seed field is improved but
its strength is diluted.

The above show that a network of wiggly cosmic superstrings may well be 
responsible for the galactic magnetic fields even though $G\mu$ is small enough
not to dominate structure formation.

\subsubsection{Superconducting strings}

Another realisation of a PMF generation over inter-string distances was 
investigated by one of us (KD) considering a network of superconducting cosmic 
strings \cite{mine}. Since D-strings arise in supersymmetric theories then
they will have fermion zero modes in the string core \cite{DDT2}. It was shown
in \cite{DDT2} that some zero modes survive supersymmetry breaking, depending
on the details of the breaking mechanisms. Consequently,
cosmic superstrings can also develop currents. If the current is 
electromagnetically coupled, then they could be superconducting. 

As in wiggly strings, superconducting strings have 
different $\tilde\mu$ and $T$. Hence, they too generate an attractive 
gravitational field. Indeed, as shown in Ref.~\cite{mine}, in this case we have
\begin{equation}
h_{00}=-4G[J^2+(\tilde\mu-T)]\ln(r/r_0)-4GJ^2[\ln(r/r_0)]^2,
\label{h001}
\end{equation}
where $J$ is the string current and
\begin{equation}
\tilde\mu\simeq\mu+\frac{J^2}{4Ke^2}\qquad{\rm and}\qquad
T\simeq\mu-\frac{J^2}{4Ke^2}
\end{equation}
with $e$ being the charge of the current carriers and \mbox{$K\gsim 1$} is a 
constant depending on the underlying model. Using the above in Eq.~(\ref{f}) 
we obtain the attractive force per unit length
\begin{equation}
f=-\frac{2GJ^2}{r}\left[1+\frac{\tilde\mu-T}{J^2}+2\ln(r/r_0)\right]\,.
\end{equation}
The deficit angle this time is given by \cite{mine}
%
\begin{eqnarray}
 & \Delta=8\pi G\left\{\tilde\mu+J^2
\left[\frac{1}{2}+\ln(r/r_0)\right]\right\} &
\label{D1}
\end{eqnarray}
Hence, the total boost is
\begin{equation}
u=8\pi G\mu v_s\gamma_s+
4\pi G(QJ)^2\left(v_s\gamma_s+\frac{1}{v_s\gamma_s}\right)\,,
\label{u3}
\end{equation}
where \mbox{$Q\sim{\cal O}(10)$} is a constant (associated with the string 
radius) due to the self-inductance of the string. Using the above, 
Eq.~(\ref{Bf}) gives
\begin{equation}
B_f\sim\frac{m_p}{e}\frac{8\pi G\mu\gamma_s}{P^{\beta}t_f}
\left\{1+\frac{(QJ)^2}{2\mu}\left[1+\frac{1}{(v_s\gamma_s)^2}\right]\right\}.
\label{Bf2}
\end{equation}
If the string velocity is relativistic then the first term in the curly 
brackets dominates the right-hand-side of the above. This is because the string
current is bounded from above as
\begin{equation}
J\leq J_{\rm max}\equiv e\sqrt{\mu}\,.
\label{Jmax}
\end{equation}
Hence, if \mbox{$v_s\gamma_s\sim 1$} the generated PMF does not differ in 
strength and coherence from the wiggly string case, resulting in a seed field
with the characteristics shown in Eqs.~(\ref{Bseed1}) and (\ref{lseed1}).

However, there is a chance that one may generate a much stronger PMF.
Indeed, in an earlier work of ours, we have shown that, if the strings
carry electrically charged currents, excessive friction between the strings and
the plasma can result in strong damping of the string motion \cite{ours}. 
The reason is that a charged current carrying string is surrounded by a 
Biot-Savart magnetic shield, which encloses the string in a magnetocylinder, 
similar to the Earth's magnetosphere. The magnetocylinder is impenetrable by
the ionised plasma, which is deflected away from the path of the moving string,
resulting in a friction force, which may heavily damp the string motion. 
In Ref.~\cite{ours} we have studied the dynamics and the evolution of such a 
string network. We have found that the strings reach a terminal velocity $v_T$
given by
\begin{equation}
v_T^2\sim\frac{G\mu}{\sqrt{GJ^2}}\,,
\label{vT}
\end{equation}
which could be rather small \mbox{$v_T\ll 1$} if the string current is large 
enough. As we have shown in Ref.~\cite{ours}, the string network in this case 
does reach a scaling solution (fixed number of strings per horizon volume), 
which, however, is much denser than the usual case. In the case of cosmic 
superstrings the density of the string network would be further increased by a
factor $P^{-\beta}$ due to the small intercommutation probability. 

Using the estimate of the terminal velocity in Eq.~(\ref{vT}), it can be easily
checked that, if the string current is
\begin{equation}
J>J_*\equiv e^{-1}(G\mu)^{1/6}J_{\rm max}\;,
\label{J_*}
\end{equation}
then the last term on the right-hand-side of Eq.~(\ref{Bf2}) is the dominant 
\footnote{This is the term due to the gravitational attraction.}. 
Then the generated PMF is 
\begin{equation}
B_f\sim\frac{4\pi Q^2\sqrt{e}\,m_p}{P^{\beta}t_f}\sqrt{G\mu}
\left(\frac{J}{J_{\rm max}}\right)^3,
\label{Bf3}
\end{equation}
where we used Eqs.~(\ref{Bf2}), (\ref{Jmax}) and (\ref{vT}). Evaluating the 
above at recombination and considering the amplification of order 
${\cal O}(10^2)$ due to the gravitational collapse of the galaxy, we find
\begin{equation}
B_{\rm seed}\sim 10^{-14}P^{-\beta}\sqrt{G\mu}\,(J/J_{\rm max})^3\;{\rm Gauss}
\label{Bseed2}
\end{equation}
which can be quite sufficient for the dynamo for 
\mbox{$J_*<J\lsim J_{\rm max}$}.

Now, the coherence of this PMF is determined by the inter-string distance, i.e.
\begin{equation}
\ell_f\sim P^{\beta}v_Tt_f\sim\frac{P^{\beta}t_f}{\sqrt e}(G\mu)^{1/4}
\left(\frac{J_{\rm max}}{J}\right)^{1/2},
\label{lf2}
\end{equation}
where we used Eqs.~(\ref{Jmax}) and (\ref{vT}). Evaluating again at 
recombination and also considering that galactic gravitational collapse reduces
the coherence of the PMF by ${\cal O}(10^{-1})$, we obtain
\begin{equation}
\ell_{\rm seed}\sim 10^{-4}P^{\beta}(G\mu)^{1/4}
\sqrt{J_{\rm max}/J}
\;\;{\rm Mpc}\,,
\label{lseed2}
\end{equation}
which only marginally satisfies the bound in Eq.~(\ref{lbound}).

Hence, from the above, we see that superconducting strings can result in 
strong and coherent magnetic fields which can trigger the galactic 
dynamo and explain thereby the observed galactic magnetic fields.

\section{Conclusions}

We have investigated the generation of a Primordial Magnetic Field (PMF) in
the Universe, through the effect of a network of cosmic superstrings onto
ionised plasma, after recombination. The PMF is created by the Harrison--Rees
mechanism, which considers a spinning volume of ionised plasma, in which 
the electron and ion fluids are spinning with different angular velocity.
Hence, circular currents arise that give birth to a PMF. The network of 
cosmic superstrings can cause such vortical motions in the plasma, due to
the gravitational effects of moving string segments. There are two 
possibilities; one being vortical motions inside string wakes, caused by the 
rapidly oscillating wiggles on the strings, and the other being vortical
motions over inter-string distances caused by the relative motion of
travelling neighbouring strings of the network. The latter effect can be
further intensified if the strings exert an attractive gravitational field
onto surrounding matter, as is the case with wiggly or superconducting strings.
In our work we have focused on the possibility that such a PMF can be 
sufficiently strong and coherent to seed the galactic dynamo mechanism and 
explain the observed galactic magnetic fields.

We have studied all the above cases and found that it is always possible to
create a PMF strong enough to seed the galactic dynamo. However, achieving 
the required coherence for this field is more challenging. Indeed, for PMF 
generation due to string wiggles, we have found that a seed field as coherent 
as \mbox{$\ell_{\rm seed}\sim 100$ pc} can be created only if the string 
tension assumes its highest possible value \mbox{$G\mu\sim 10^{-7}$}. This 
constraint is somewhat relaxed if the PMF is generated as late as the late 
reoinisation period \mbox{$z_{\rm ri}\sim 15$}, but, in this case, only a 
fraction of the galaxies is expected to be magnetised, because the string wakes
are far apart. In the case of inter-string PMF generation, coherence is easier 
to attain. Indeed, a coherent enough seed field can be obtained provided the 
intercommutation probability is not extremely low. Cosmic superstrings that
carry substantial currents may generate a really intense PMF (up to 
\mbox{$B_{\rm seed}\sim 10^{-15}$Gauss}) but at the expense of its coherence.
An adequately coherent field, in this case, also requires a high value for 
$G\mu$.

Cosmic superstrings are a probable result of brane cosmology and brane 
inflation models. However, since they are not the primary cause of
the acoustic peaks in CMB spectrum, such strings appeared to have 
limited observational signatures. As a result, their observable 
consequences were thought to be limited to gravitational lensing events. 
In this paper we show that cosmic superstrings may have further important 
cosmological consequences. In particular, they may be responsible for the 
observed galactic magnetic fields and, in general, long range magnetic 
fields in the IGM. Since the stellar magnetic dynamo is thought to be 
triggered by the galactic magnetic field, it seems plausible that cosmic 
superstrings could be the principle source of magnetisation in the Universe.


\end{document}